\documentclass[preprint]{aastex}
\usepackage{natbib,psfig,emulateapj5}
\bibpunct{(}{)}{;}{a}{}{,}

\newcommand{\CIV}{\ion{C}{4}}
\newcommand{\MgII}{\ion{Mg}{2}}

\begin{document}

\received{}
\accepted{}
%

\title{AGN Outflows in Emission and Absorption: The SDSS Perspective\thanks{}}

\author{
Gordon T. Richards\altaffilmark{2}
}
\altaffiltext{1}{Summary of talk presented at the ``AGN Winds in the Caribbean'' Workshop, St.\ John, USVI; 28 November -- 2 December, 2005; http://www.nhn.ou.edu/$\sim$leighly/VImeeting/}
\altaffiltext{2}{Department of Physics \& Astronomy, Johns Hopkins University, 3400 N. Charles St., Baltimore, MD 21218.}

\begin{abstract} 

A variety of investigations have demonstrated commonalities between
the Baldwin (1977) Effect, the blueshifting of \CIV\ emission lines
\citep[e.g.,][]{gas82,rvr+02}, and the $L_{UV}$--$L_{X}$ relationship
\citep[e.g.,][]{at82,sbs+05,ssb+06}; indeed all three of these
observational effects may be manifestations of the same underlying
(but still uncertain) physics.  This commonality is of interest to
investigations of accretion disk winds \citep[e.g.,][]{mcg+95,psk00}
from active galactic nuclei (AGN) as there is evidence that broad
absorption line quasars (BALQSOs) are drawn from a parent sample of
quasars that exhibit larger than average \CIV\ blueshifts, weaker than
average \CIV\ emission line strengths, and bluer than average
(intrinsic) colors.  The properties of the absorption troughs appear
to be dependent upon these parameters.  Thus, it is suggested that not
all quasars will host bona-fide BAL troughs, but that all (broad
emission line) quasars host outflows of some type, the structure of
which is strongly dependent on the quasar's spectral energy
distribution.

\end{abstract}

\section{Introduction/Review of \citet{rvr+02}}

While quasar spectra appear quite similar, there are a number of
well-studied distinctions that provide a basis for quasar
sub-classification.  Two of those effects involve changes in the \CIV\
emission line; these are the \citep{bal77} Effect (BEff) and the
blueshifting of the peak of the \CIV\ emission line with respect to
the quasars' systemic redshift \citep{gas82,wil84}.  In the interim
years, numerous papers have considered these phenomena, exploring
their dependence on radio loudness, luminosity, the shape of the
spectral energy distribution and the extent to which lines other than
\CIV\ are effected
\citep[e.g.,][]{mf84,kk86,bps+89,tf92,zm93,bws+94,msd+96,kbf98,elv00,smd00,lm04,lei04,wvb+06}

Recently, \citet[][hereafter R02]{rvr+02} investigated the
blueshifting of \CIV\ emission for a large sample ($>700$ quasars)
drawn from the Sloan Digital Sky Survey (SDSS; \citealt{yaa+00}).  R02
confirmed previous results that showed that \CIV\ blueshifts are
ubiquitous in radio-quiet quasars and that radio-loud quasars tend
towards smaller blueshifts; see also \citet{msd+96}.  It was also
shown that quasars with large blueshifts have \CIV\ emission lines
that have larger FWHM, but smaller equivalent widths than the average
quasar; see also \citet{cor90}.  R02 found that controlling for
luminosity does not eliminate the effects of \CIV\ blueshifts, but
that large blueshift quasars are somewhat bluer and more luminous than
quasars with small blueshifts.

An outflow such as from an accretion disk wind
\citep[e.g.,][]{mcg+95,mc97,psk00,cn03,eve05} is often cited as the
most plausible explanation for the origin of \CIV\ blueshifts
\citep[e.g.,][]{lei01}.  While an outflow related origin for this
effect seems likely, R02 pointed out that there is good agreement in
the blue wing between blueshifted and un-blueshifted objects and
argued that these blueshifts are not well explained by a simple
translation of the line, but rather that the red wing is
preferentially affected by absorption, suppression, and/or
obscuration; see also \citet{lei01}.  It is important to consider this
possibility as it would have strong negative implications for the use
of the \CIV\ emission line in black hole mass (and possibly
metallicity) determinations.

Finally, R02 suggested that \CIV\ blueshifts may represent an
orientation effect.  This could be in terms of an external orientation
effect such as the tilt of the accretion disk with respect to our line
of sight, or it could be an internal orientation effect such as the
opening angle of the accretion disk wind (e.g.,
\citealt{elv00}, Fig.~7; \citealt{ekk00}, Figs.~1 and 2).  Current
evidence would seem to favor the latter interpretation as radio-loud
objects preferentially exhibit small blueshifts --- regardless of
their radio morphology (and thus orientation).

Finally, it is important to emphasize that the \CIV\ blueshift effect
is related to the results derived from eigenvector decomposition
\citep{bg92,bf99,smd00,bor02,ycv+04} and that it has been argued
\citep[e.g.,][]{bl05} that the \CIV\ line profile is governed by
$L/L_{Edd}$.


\section{New/Updated Blueshift Results}

\begin{figure}
\plotone{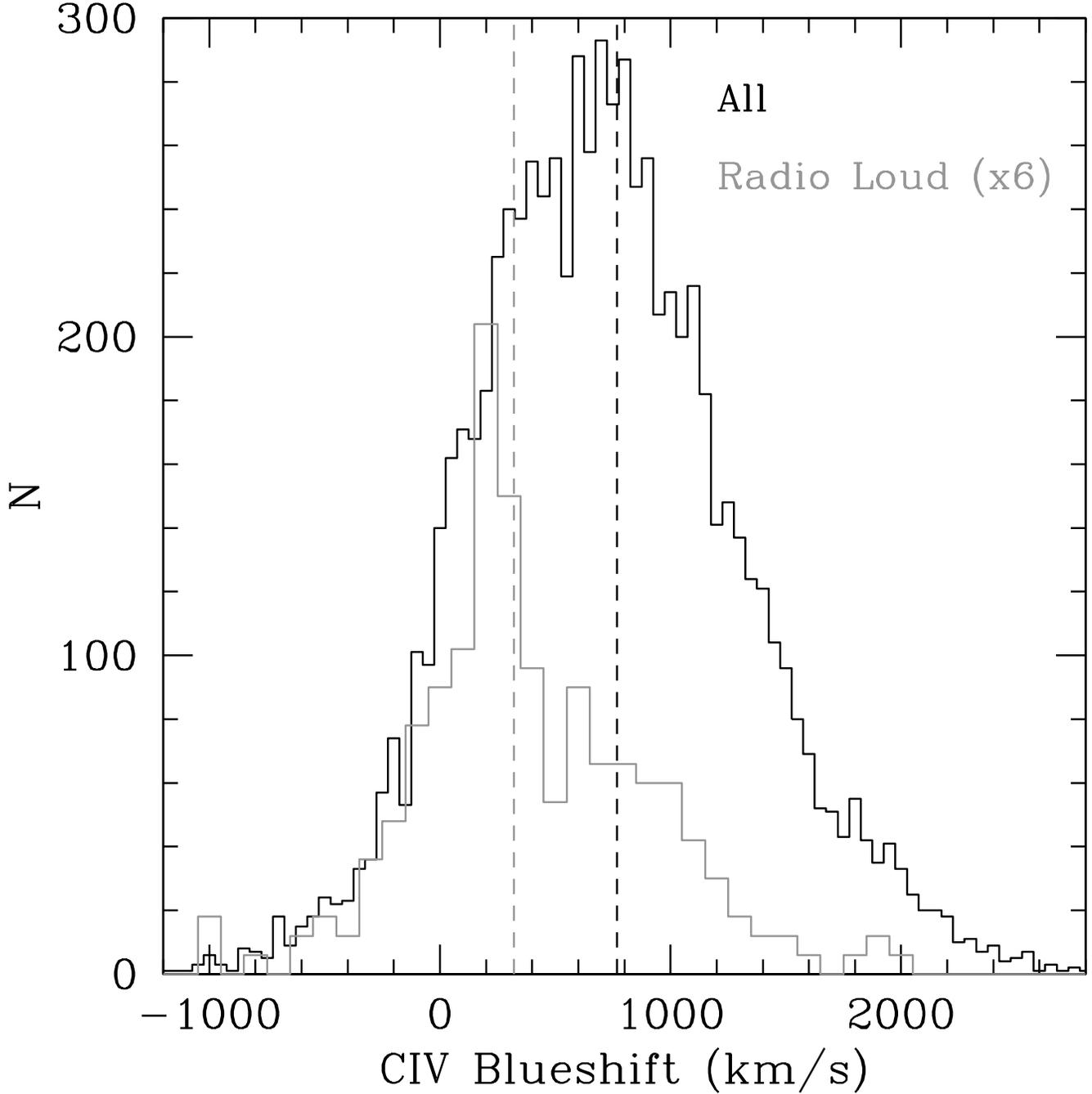}
\caption{Blueshift of the CIV emission line (with respect to MgII)
for over 7500 SDSS-DR4 quasars \citep{aaa+06}.  Positive values
indicate a blueshift of CIV (and thus erroneously small systemic
redshift) with respect to MgII.  Only non-BAL quasars with small
errors ($<300$ km/s) in this quantity are shown. Radio-loud quasars
($\log L_{rad}>33$ [ergs/s/Hz]) are shown in gray after scaling by a
factor of 6 for comparison with the full sample.  Dashed lines
indicate the median values.
\label{fig:fig1}
}
\end{figure}

With a much larger sample of quasars than the SDSS Early Data Release
sample used by R02, it is now possible to explore trends that were not
possible to explore before.  Figure~\ref{fig:fig1} shows the
distribution of \CIV\ blueshifts for over 7500 SDSS-DR4 \citep{aaa+06}
quasars --- nearly 10 times the sample size from R02.  It is quite
clear that these shifts of the \CIV\ emission line with respect to
\MgII\ are ubiquitous for the population as a whole, yet are much
smaller for radio-loud quasars.

\begin{figure}
\plotone{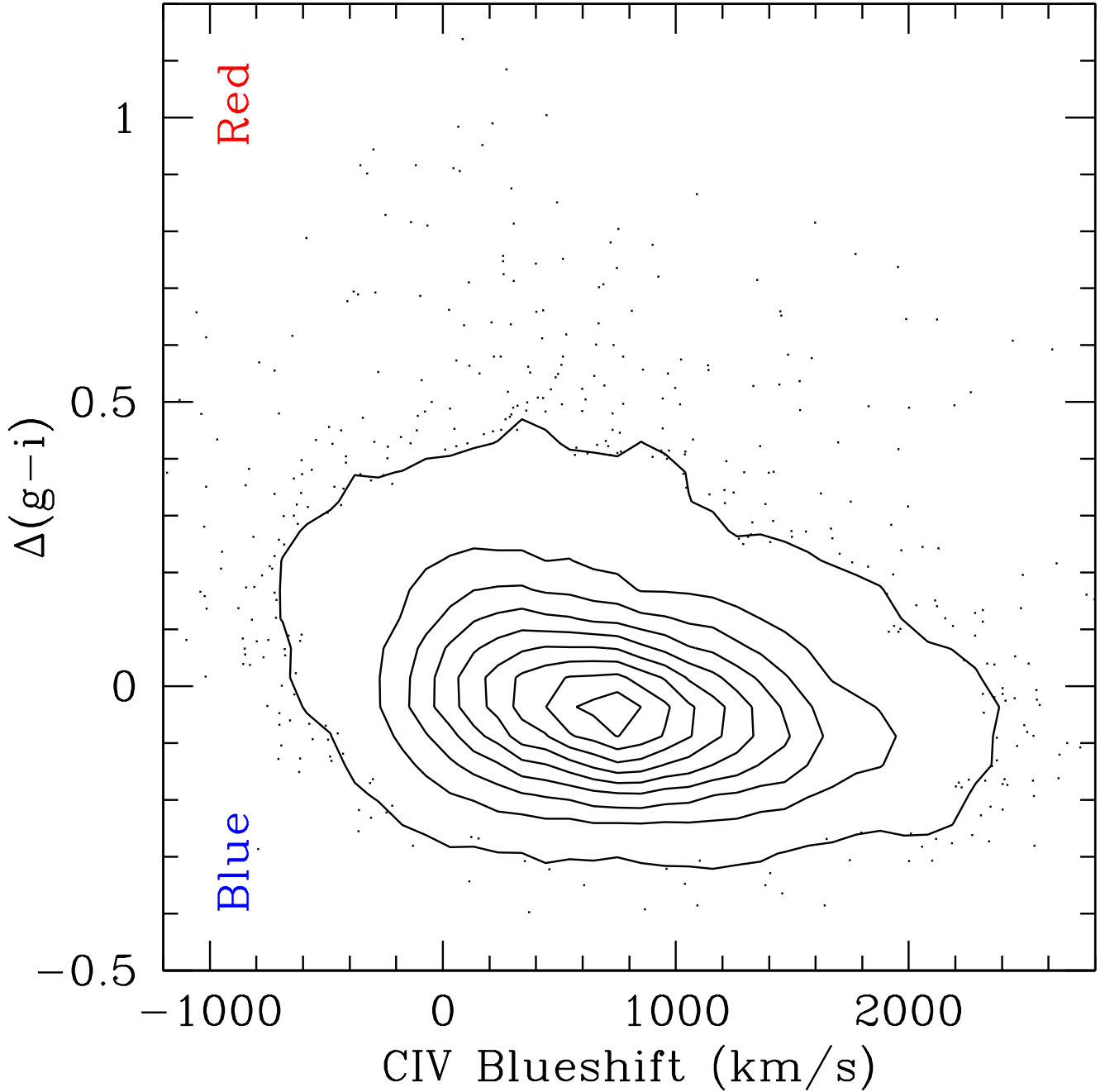}
\caption{Blueshift of the CIV emission line versus relative $g-i$
colors, $\Delta(g-i)$.  The relative color is computed by subtracting
the mean $g-i$ color as a function of redshift from each observed
value and is a robust indicator of optical spectral index that is
independent of redshift and (average) emission line strength
\citep{rhv+03}.  R02 constructed composites in 4 bins of CIV
blueshift, but it is clear that there are differences in optical color
that should be accounted for.  While quasars with large CIV blueshift
are generally blue, not all blue quasars have large CIV blueshifts.
\label{fig:fig2}
}
\end{figure}

\begin{figure}
\plotone{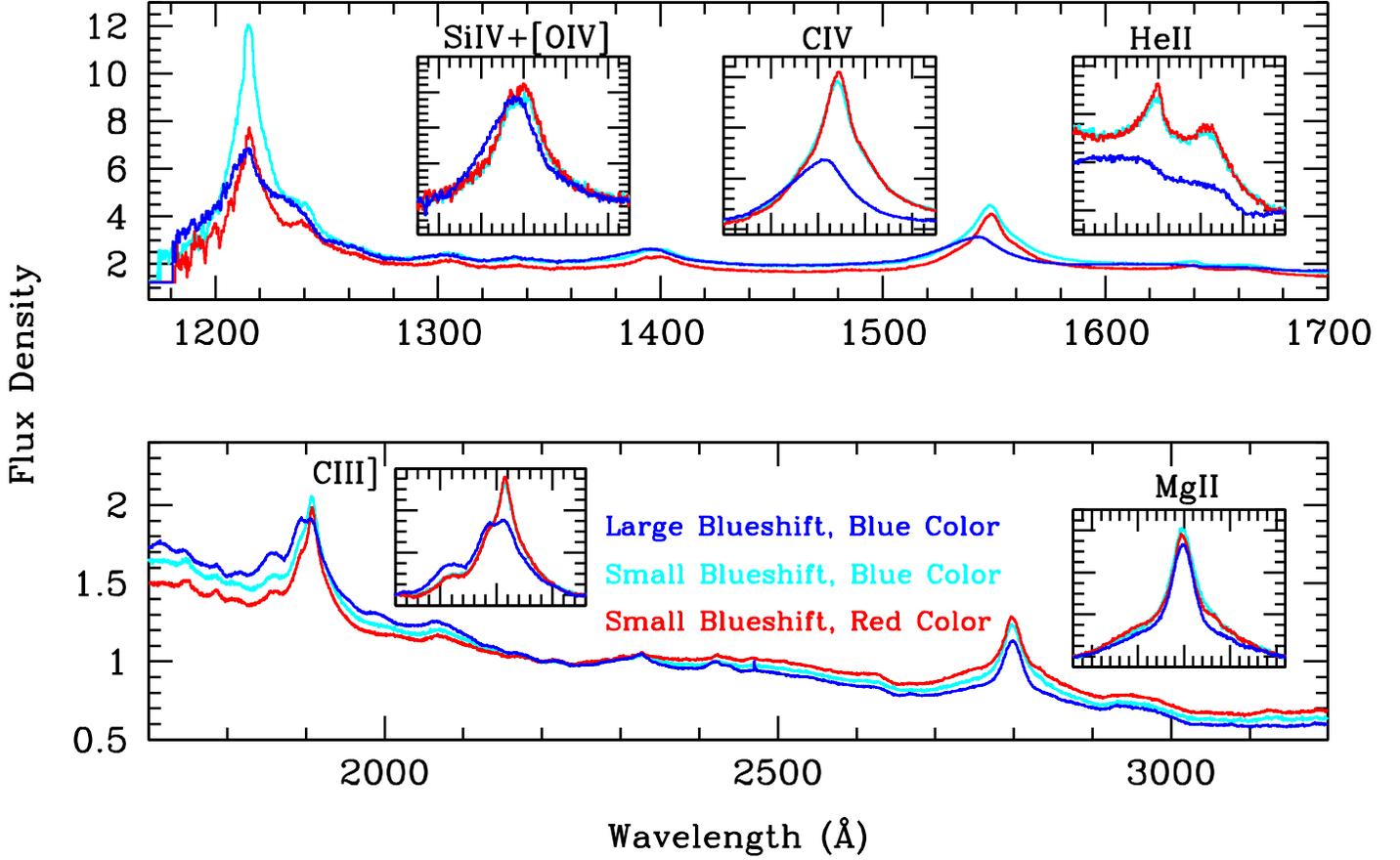}
\caption{Composite spectra of blue, large blueshifted (blue); blue,
small blueshifted (cyan); and red, small blueshifted (red) quasars.
Each composite contains over 600 quasars.  The main panels are
normalized at 2200 \AA, for comparison of the spectral slopes, while
the insets have local normalizations so that the line profiles can be
compared.  Note that, with the exception of Ly$\alpha$, the blue and
red composites with small CIV blueshifts have very similar emission
line features.  Likely dust reddened quasars [$\Delta(g-i)>0.2$] have
been excluded in all composites.
\label{fig:fig3}
}
\end{figure}

R02 built composite spectra in 4 bins of \CIV\ blueshift and showed
that, on the average, quasars with large \CIV\ blueshifts have bluer
than average continua.  Similarly, \citet{rhv+03} showed that bluer
quasar tend to have weaker emission lines (particularly \CIV).
However, further work has found that there is not a simple one-to-one
relationship between the continuum color and the emission line
strength/blueshift \citep[e.g.,][]{grh+05}.  This can be seen in
Figure~\ref{fig:fig2} which shows that while quasars with large
\CIV\ blueshifts are generally blue, not all blue quasars have large
\CIV\ blueshifts.  Thus it is necessary to reconsider the composite
spectra properties investigated by R02 as a function of color in
addition to blueshift.  Figure~\ref{fig:fig3} depicts the results
of making composite spectra of quasars drawn from three extrema in
Figure~\ref{fig:fig2}: blue quasars with large \CIV\ blueshifts,
blue quasars with small \CIV\ blueshifts, and red quasars with small
\CIV\ blueshifts.  The samples are designed to have roughly equal
numbers in each category, with over 600 quasars in each bin.  Red and
blue quasars with small blueshifts have quite similar emission
features (perhaps with the exception of Ly$\alpha$).  On the other
hand, blue quasars with large and small blueshifts have drastically
different emission line properties despite quite similar continua.

\begin{figure}
\plotone{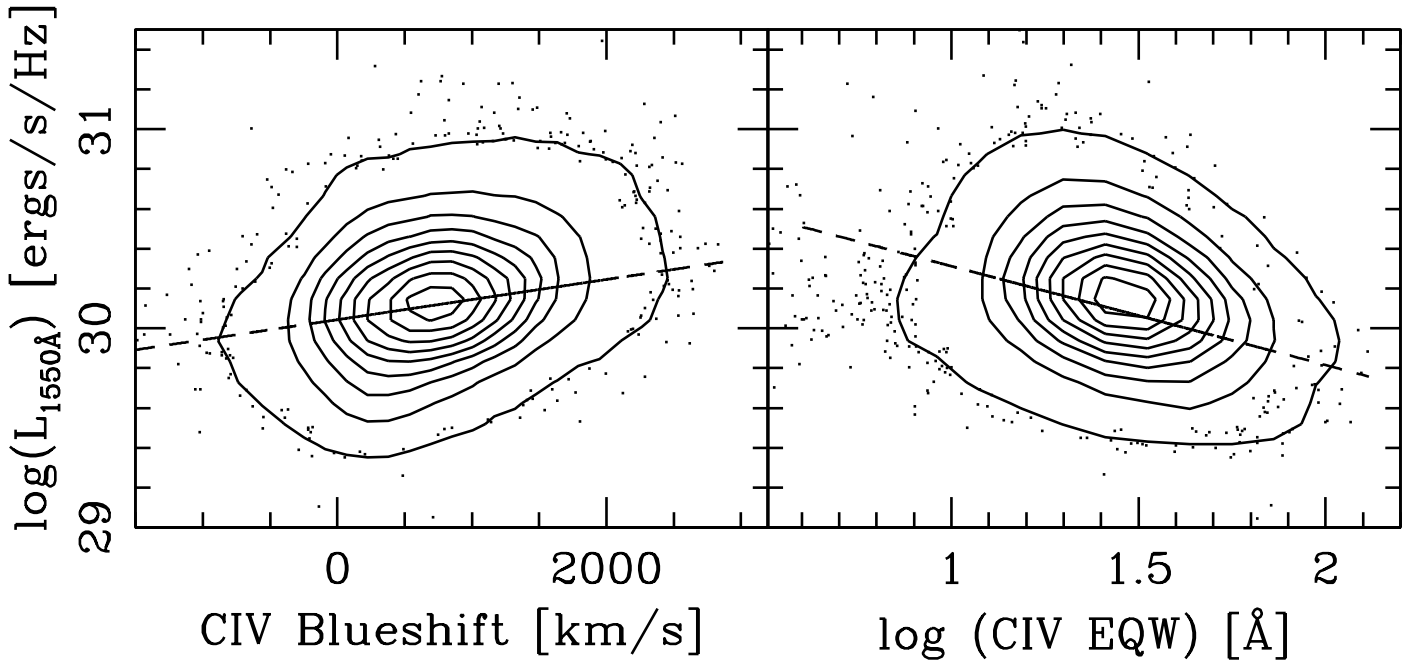}
\caption{There are weak trends between UV luminosity ($L_{1550{\rm\AA}}$)
and CIV blueshift and CIV equivalent width.  The latter being the
Baldwin (1977) Effect.
\label{fig:fig4}
}
\end{figure}

An obvious question is whether these differences are due to
luminosity.  R02 showed that, while quasars with large blueshifts are
somewhat more luminous and bluer, controlling for luminosity effects
(using $M_i$) did not mitigate the \CIV\ blueshift effect.
Nevertheless, it is still true that there exists a weak trend with
luminosity, in a manner similar to the Baldwin Effect, see
Figure~\ref{fig:fig4}.

\begin{figure}
\plotone{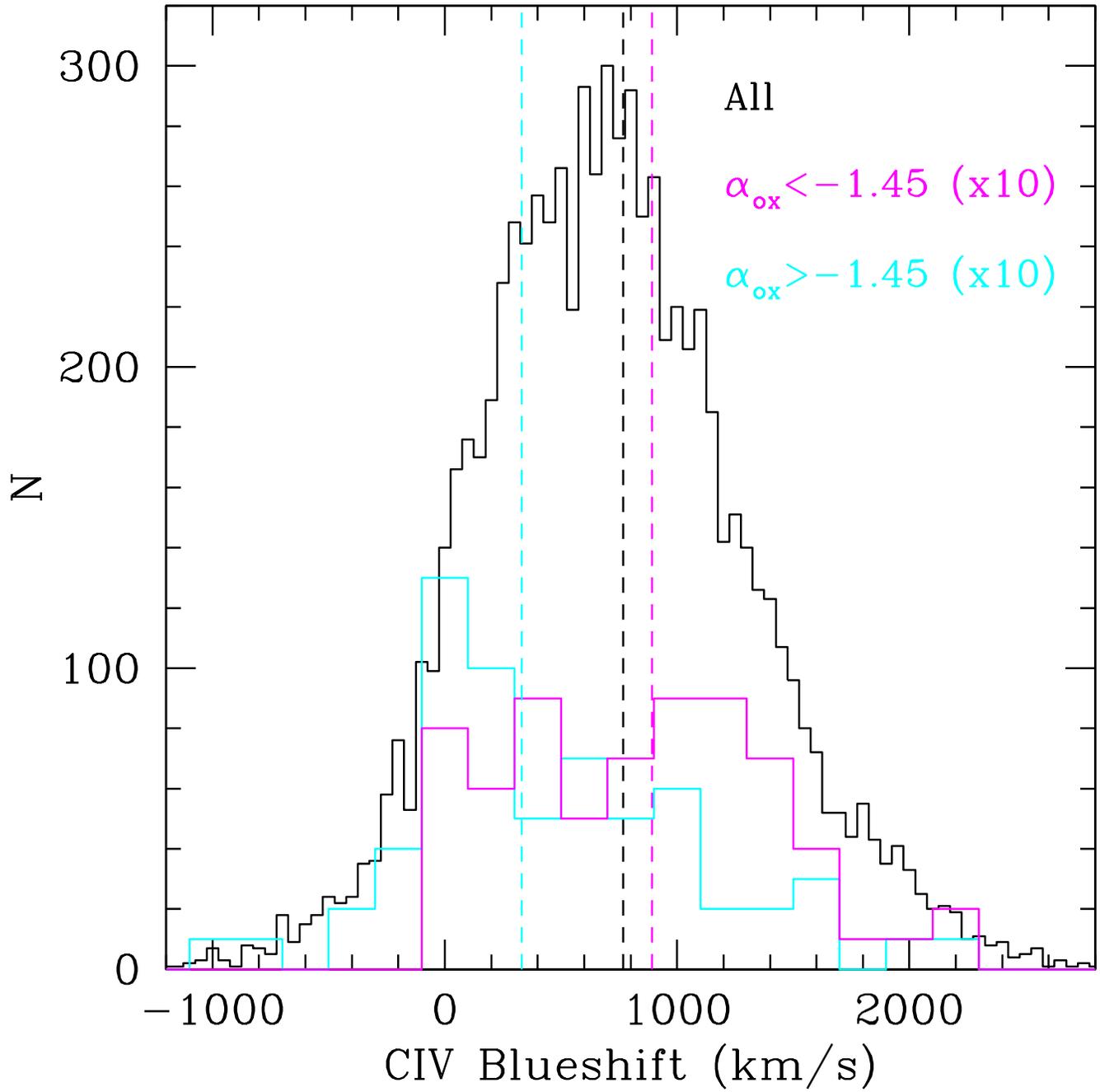}
\caption{There is a weak tendency for quasars that are X-ray weak
(relative to the UV/optical) to have larger blueshifts.
\label{fig:fig5}
}
\end{figure}

R02 also investigated the blueshift distribution as a function of
radio and X-ray brightness.  It was found that radio-loud quasars had
smaller blueshifts as is demonstrated in Figure~\ref{fig:fig1}.
X-ray (ROSAT) detected objects, however, showed a more even
distribution across all blueshifts, but with a slight tendency towards
X-ray detections being associated with smaller blueshifts
(particularly for pointed ROSAT observations).  Here we have a large
enough sample to extend this investigation.  Using the
(pipeline-measured) 2800 \AA\ continuum flux density (at \MgII) and
the ROSAT count rate we compute $\alpha_{ox}$, the ratio of flux
densities at 2500\AA\ and 2keV comparison (assuming
$\alpha_{\nu}=-0.5$).  As can be seen in Figure~\ref{fig:fig5},
X-ray strong quasars ($\alpha_{ox}>-1.45$) have smaller than average
blueshifts --- perhaps consistent with the X-ray strength of
radio-loud quasars \citep{swe+93}.
On the other hand, X-ray weak quasars tend towards larger blueshifts.
However, as $\alpha_{ox}$ is correlated with $L_{\rm UV}$ \citep[][and
references therein]{ssb+06}, it is difficult to determine whether this
trend is due to the optical or X-ray.

Along these lines, it is interesting to note that using {\em Chandra}
data for a dozen SDSS quasars representing the extrema in the \CIV\
blueshift distribution, \citet{grh+05} showed that large blueshift
objects have somewhat steeper hard X-ray spectra than small blueshift
quasars and that the large blueshift quasars appear to be absorbed in
soft X-rays.  This trend is consistent with our finding here that
large blueshift quasars are more X-ray weak.  

Thus, large \CIV\ blueshift quasars may be more UV luminous than
average, have bluer than average UV spectral indices, and have steeper
than average hard X-ray spectra (that are either absorbed or
intrinsically weak in soft X-rays).  These are the very properties
needed for effective radiation pressure driving of an accretion disk
wind \citep[e.g.,][]{psk00,lei04}.

\section{\CIV\ Blueshifts $=$ the Baldwin Effect}

It has long been realized that the $L_{UV}$--$L_X$ relationship and
the BEff may not be independent \citep[e.g.,][]{zm93,gre98,kbf98},
particularly given that both relationships depend on the optical/UV
luminosity and that the soft X-ray spectrum (47.9--64.5 eV) is
responsible for the relative abundance of triply ionized carbon.

\begin{figure}
\plotone{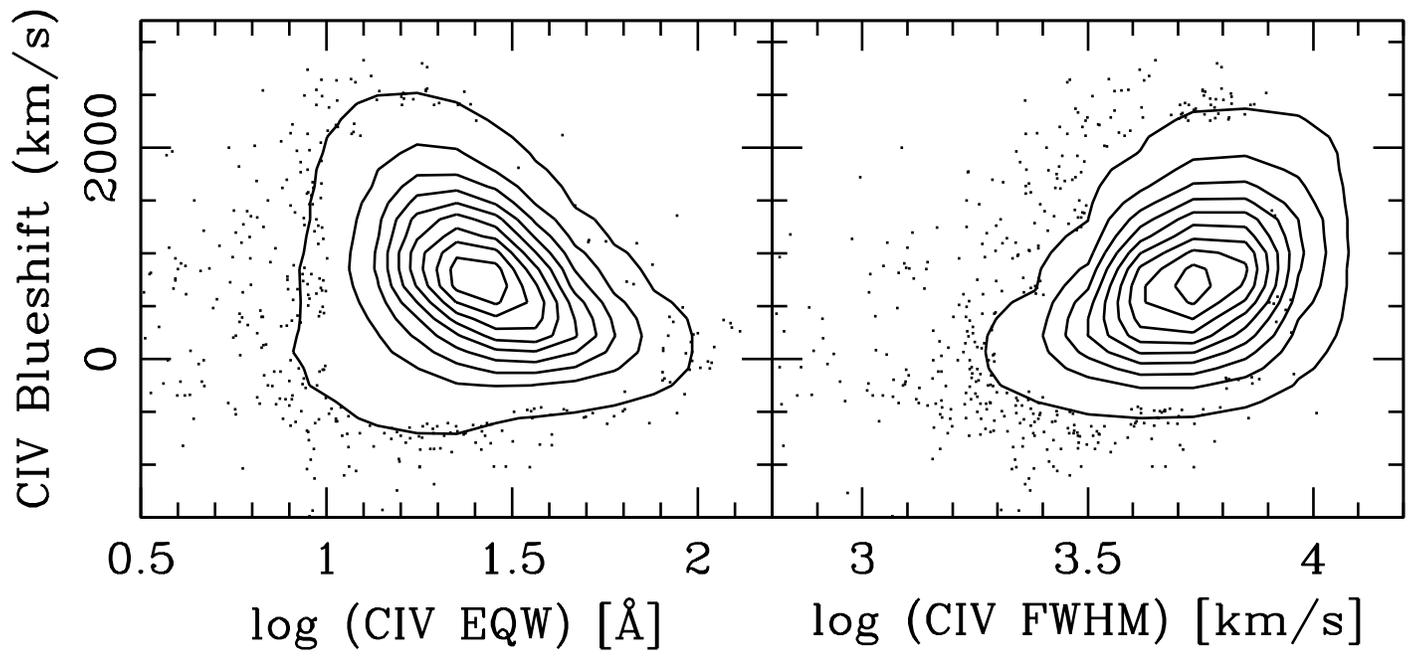}
\caption{Quasars with large blueshifts have small equivalent widths and ``broader'' lines.
\label{fig:fig6}
}
\end{figure}

Here that argument is extended to include \CIV\ blueshifts.  That is,
\CIV\ blueshifts, the BEff, and the $L_{UV}$--$L_X$ relationship are
suggested as having common origin; see also \citet{lei04}.  The
arguments for this commonality are simple, yet powerful.  Two primary
lines of evidence suggest a unification of the BEff and the
blueshifting of \CIV\ emission lines.  First, as with the BEff, there
is a weak luminosity trend in the \CIV\ blueshifts: quasars with large
blueshifts are slightly more luminous, see the left hand panel of
Figure~\ref{fig:fig4}.  Second, quasars with large blueshifts
have smaller \CIV\ equivalent widths, see the left hand panel of
Figure~\ref{fig:fig6}. Thus, $L_{UV}$, \CIV\ equivalent width,
and \CIV\ blueshift appear to be correlated with each other.  None of
these trends are particularly tight --- there is large scatter in the
luminosity trends and any correlation between \CIV\ blueshift and
\CIV\ EQW are largely due to a lack of large EQW lines in large
blueshift objects; however, the trends do appear to be robust.

A connection between \CIV\ blueshifts and the $L_{UV}$--$L_X$
relationship is supported on both an empirical and a theoretical
basis.  Empirically, we saw in Figure~\ref{fig:fig5} that
quasars that are stronger in the X-ray relative to the optical/UV
($\alpha_{ox}$ less negative) have smaller \CIV\ blueshifts.
Theoretically, it is perhaps not surprising that changes to the \CIV\
emission line are seen with increasing UV luminosity (and thus
[relatively] less X-ray luminosity) given that the production of \CIV\
is regulated by the incidence of X-ray photons.

Further supporting these connections is the similarity of the emission
lines involved in each effect (and to what extent) and their
ionization potential.  For, example, \ion{He}{2} 1640\AA\ has a trend
that is highly correlated with the \CIV\ blueshift, and, as emphasized
by \citet{lm04}, weak \ion{He}{2} may be indicative of a relative
paucity of soft X-rays.  Also, \ion{Si}{4} (which may have a strong
contribution from \ion{O}{4}]) shows a weak blueshift effect at best
(see the insets in Figure~\ref{fig:fig3}).  These trends are
consistent with the general trends for the BEff
\citep[e.g.,][]{dhs+02}.  Finally, while not all previous work agrees
on this point, \citet{fk95} showed that the Baldwin Effect is
strongest in the red wing of the \CIV\ emission line.  This is
equivalent to saying that \CIV\ blueshifts and the BEff are the same
phenomenon.

Thus, both the BEff and \CIV\ blueshifts are seen under similar
conditions, produce similar effects, and have the same dependence on
the X-ray part of the spectrum, suggesting a commonality between these
relationships.  The equality of {\em all three} of these effects is
not generally appreciated in part because the \CIV\ emission line is
often used to define a quasar's systemic redshift (thus obscuring the
blueshift effect) and because it can be difficult to disentangle SED
effects from luminosity effects (especially when using broad-band
luminosities).

\section{The BALQSO Parent Sample}

\begin{figure}
\plotone{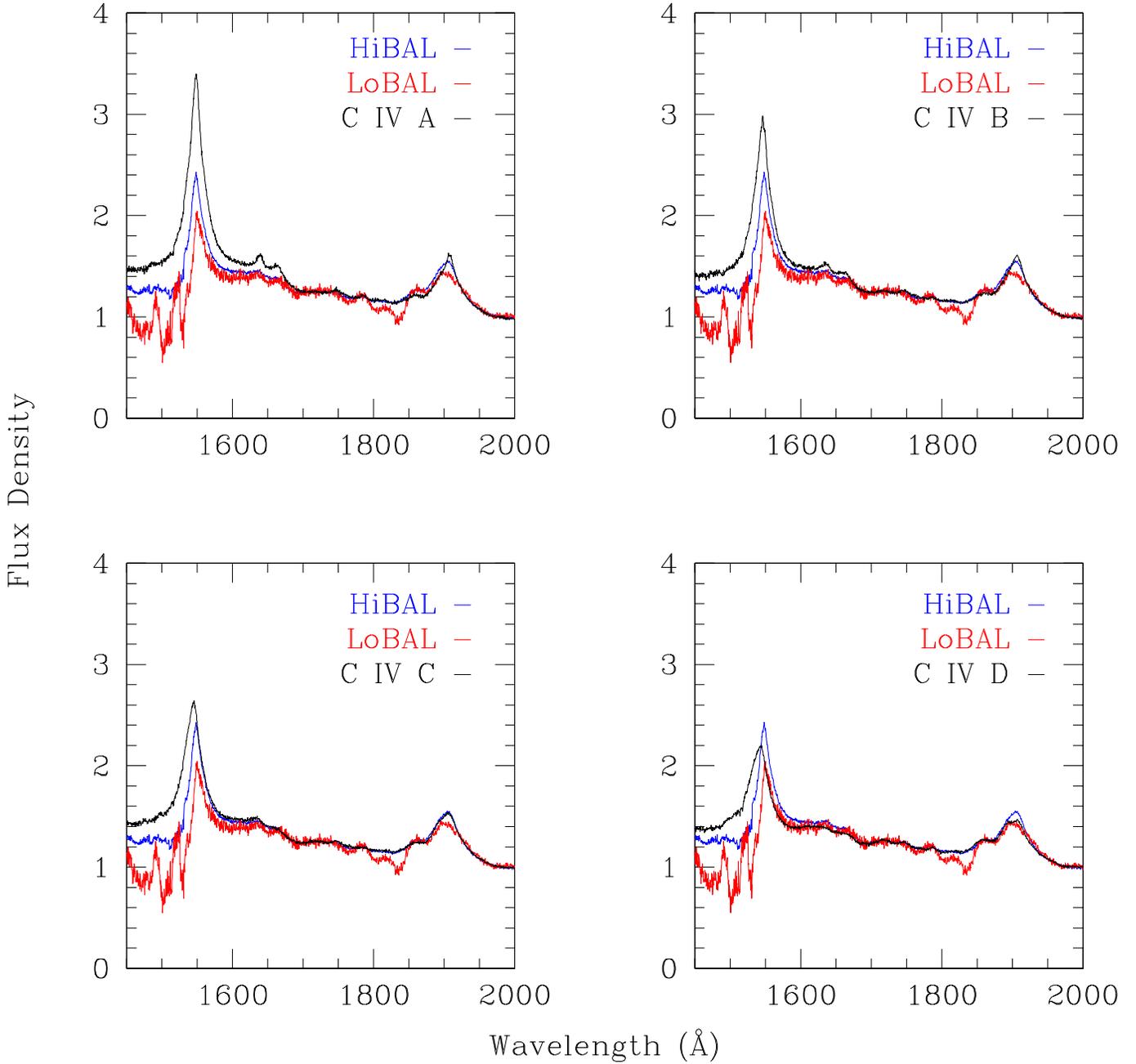}
\caption{Comparison of HiBAL and LoBAL composites to CIV blueshift
composites.  Blueshift composite A (upper left hand panel) has the
smallest CIV blueshifts, composite D (lower right hand panel) has
the largest blueshifts.  By comparison with the red wing of CIV and
the CIII] emission line region, it is seen that BALQSOs have
larger than average CIV blueshifts, LoBALs being more extreme than
HiBALs.  (Adapted from R02).
\label{fig:fig7}
}
\end{figure}

The above X-ray/blueshift relationship is particularly interesting
when considering the claims by \citet{cor90}, R02, and \citet{rrh+03}
that BALQSOs have larger than average blueshifts.  Furthermore,
BALQSOs are known to exhibit strong X-ray absorption and very negative
values of $\alpha_{ox}$ \citep[e.g.,][]{blw00,gbc+06}.  In
Figure~\ref{fig:fig7}, we show how, despite absorption blueward of
the peak of \CIV\ emission, it is still possible to use the red wing
of \CIV, the \ion{He}{2} emission line, and the \ion{C}{3}] emission complex to
demonstrate that BALQSOs have larger than average blueshifts (LoBALs
exhibiting the largest blueshifts).

\citet{rrh+03} and \citet{thr+06} further showed that, after
correction for dust reddening, intrinsically red and intrinsically
blue HiBALs have distinct emission and absorption properties.  ``Red''
HiBALs have much stronger \CIV, \ion{He}{2}, and \ion{C}{3}] emission
lines than ``blue'' HiBALs --- consistent with the emission line
differences seen in non-BALQSOs by \citet{rhv+03} and in
Figure~\ref{fig:fig3} above.  Further, blue HiBALs appear to
exhibit BAL troughs that extend to higher maximum outflow velocities.
These relationships are of great interest for disk-wind models,
especially when considered in concert with the results of
\citet[][Fig.~5]{gbc+06} showing that X-ray faint quasars have
weaker \CIV\ emission lines, and \CIV\ absorption troughs that are
broader and extend to higher velocities.

Thus, not only does it seem that quasars with certain SEDs (and
emission line properties) are more likely to host BALQSOs, the SED
determines what {\em kind} of BALQSO a quasar can host.  Since the
blueshift and color distributions appear continuous, we speculate that
this SED dependence extends from BALQSOs \citep{wmf+91,rrs+03,thr+06}
to those with NALs \citep{fwp+86,hbj+97,ric01,gbc+01,ves03} and from
radio-quiet to radio-loud objects
\citep[e.g.,][]{smw+92,rwg+00,mvi+01}, perhaps as a result of a
gradual transition from a radiation-dominated wind \citep{psk00} to an
MHD-dominated wind \citep{kk94,eve05}; see also \citet{ekk00},
\citet{pro03} and \citet{rhr+04}.

In this picture the 15\% BALQSO fraction
\citep[e.g.,][]{wmf+91,tkt02,rrh+03,hf03,thr+06} represents {\em neither} a
separate class of objects {\em nor} the BAL covering fraction of a
single population of objects.  But rather it is likely that the BAL
covering fraction is quite large for one extrema of the quasar
population (blue, blueshifted quasars) and is essentially zero for the
opposite extrema (though these objects likely still have outflows, it
is just that they are not strong enough to be classified as bona-fide
BALQSOs).  In this case the 15\% fraction is merely the SED-averaged
BALQSO fraction.

\begin{figure}
\plotone{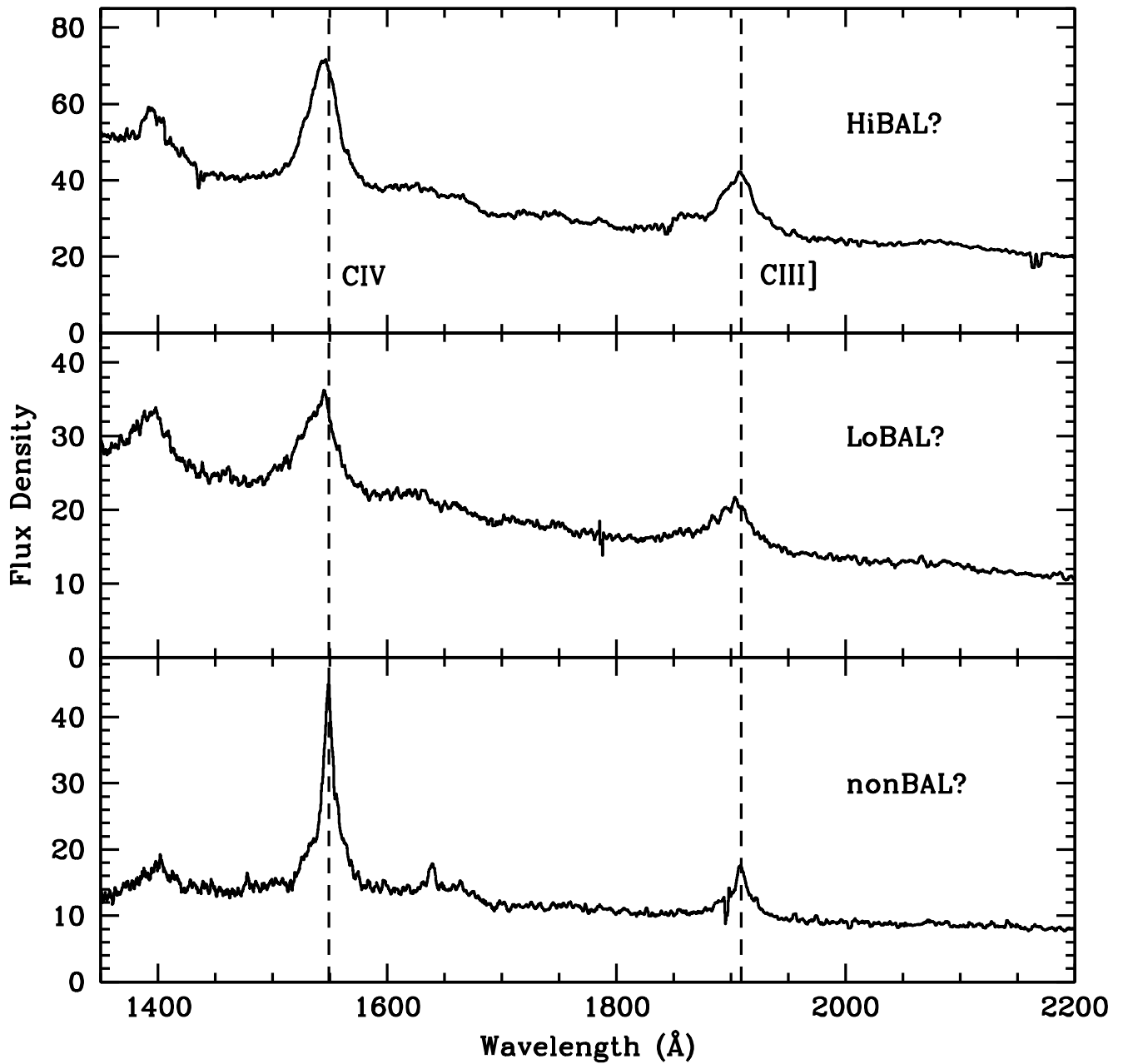}
\caption{{\em Top:} Quasar without BAL absorption along our line of
sight, but with spectral features (blue continuum, weak, blueshifted
emission lines) indicating that it is likely a BALQSO along other
lines of sight.  {\em Middle:} A more extreme version of the top
object, possibly possessing a strong enough radiation driven wind to
produce a low-ionization BAL troughs along some lines of sight.  {\em
Bottom:} A quasar with a relatively red continuum and strong CIV
emission that is not blueshifted with respect to MgII.  Such a quasar
is unlikely to exhibit BAL troughs along {\em any} line of sight,
though it may still exhibit weaker outflow features.  In each panel,
the emission line peaks are marked by dashed lines according to the
redshift derived from the MgII emission line.
\label{fig:fig8}
}
\end{figure}

We can summarize this graphically in Figure~\ref{fig:fig8}.  Blue
quasars with weak, blueshifted CIV (top panel) are the parent sample
of BALQSOs.  Along some lines of sight the top quasar will exhibit BAL
troughs.  Even bluer quasars with even weaker, more strongly
blueshifted CIV such as the middle quasar are likely to show
Low-ionization BAL troughs.  Such an object may only be a HiBAL along
some lines of sight (whereas the top quasar may not show
low-ionization troughs along any lines of sight).  Red quasars with
strong CIV that is not blueshifted, such as the bottom quasar, have
considerably different SEDs that result in significantly different
wind structures, and along no lines of sight would the bottom quasar
exhibit BAL troughs.  However, such quasars likely still have a wind
and may show intrinsic NAL absorption.  Note also that the bottom
quasar has a much greater than 10\% chance of being radio-loud, while
the top objects have little (if any) chance of being radio loud ---
the wind structure also being related to that aspect of quasar
properties.

\section{Discussion/Conclusions}

An obvious chicken and egg question underlies all of these results.
Specifically, is there a single common quasar SED whose appearance is
affected by material along our line of sight, or is there a range of
intrinsic quasar SEDs that affect the structure of the disk wind, and
thus the emission/absorption material that is seen along our line of
sight?  An argument against a common SED is that fact that radio-loud
quasars do not span the full range of SEDs or spectral properties
(e.g., emission/absorption line strength).  Even among radio-quiet
quasars the fact that the bluest quasars exhibit the greatest tendency
for UV and X-ray absorption suggests a heterogeneous intrinsic SED
distribution.  Finally, it should be remembered that disk-wind models
are highly dependent on the {\em intrinsic} UV to X-ray flux ratio
\citep[e.g.,][]{pk04,pro05}.

\acknowledgements

I thank Patrick Hall, Jonathan Trump, Timothy Reichard, and Sarah
Gallagher for the contributions that they made to this work; Karen
Leighly and the conference organizers for hosting an enjoyable and
productive meeting; and a Gordon and Betty Moore Fellowship in Data
Intensive Science for supporting this work.

Funding for the SDSS and SDSS-II has been provided by the Alfred
P. Sloan Foundation, the Participating Institutions, the National
Science Foundation, the U.S. Department of Energy, the National
Aeronautics and Space Administration, the Japanese Monbukagakusho, the
Max Planck Society, and the Higher Education Funding Council for
England. The SDSS Web Site is http://www.sdss.org/.

The SDSS is managed by the Astrophysical Research Consortium for the
Participating Institutions. The Participating Institutions are the
American Museum of Natural History, Astrophysical Institute Potsdam,
University of Basel, Cambridge University, Case Western Reserve
University, University of Chicago, Drexel University, Fermilab, the
Institute for Advanced Study, the Japan Participation Group, Johns
Hopkins University, the Joint Institute for Nuclear Astrophysics, the
Kavli Institute for Particle Astrophysics and Cosmology, the Korean
Scientist Group, the Chinese Academy of Sciences (LAMOST), Los Alamos
National Laboratory, the Max-Planck-Institute for Astronomy (MPIA),
the Max-Planck-Institute for Astrophysics (MPA), New Mexico State
University, Ohio State University, University of Pittsburgh,
University of Portsmouth, Princeton University, the United States
Naval Observatory, and the University of Washington.

\bibliographystyle{apj3}
\bibliography{agnwinds,sdsstech}

\end{document}